\documentclass[11pt,a4paper]{article}

\usepackage{amssymb}
\usepackage{amsmath}
\usepackage{amsthm}
\usepackage{mathrsfs}
\usepackage{booktabs}
\usepackage[textwidth=16.0cm,textheight=22.6cm]{geometry}
\usepackage{graphicx}
\usepackage{multirow}
\usepackage{tabularx}
\usepackage{xcolor}
\usepackage{xspace}

\theoremstyle{plain}
\newtheorem{theorem}{Theorem}

\newtheorem{proposition}[theorem]{Proposition}
\newtheorem{claim}[theorem]{Claim}
\theoremstyle{definition}
\newtheorem{definition}{Definition}

\newcommand{\algo}[1]{$\mathsf{#1}$}

\newcommand{\comment}[1]{}

\DeclareMathOperator{\N}{\mathbb{N}}

\newcommand{\mconnect}{\text{$^+\!$}}
\newcommand{\localtext}{LOCAL}
\newcommand{\QC}{\ensuremath{\mconnect \mathcal{Q}}\xspace}
\newcommand{\EN}{\ensuremath{\mconnect \mathcal{E}}\xspace}
\newcommand{\SE}{\ensuremath{\mconnect \mathcal{S}}\xspace}
\newcommand{\LOCAL}{\ensuremath{\mathcal{LOCAL}}\xspace}
\newcommand{\CONGEST}{\ensuremath{\mathcal{CONGEST}}\xspace}
\newcommand{\LOCALQE}{\ensuremath{\mathcal{\localtext\mconnect Q\mconnect E}}\xspace}
\newcommand{\LOCALQS}{\ensuremath{\mathcal{\localtext\mconnect Q\mconnect S}}\xspace}
\newcommand{\LOCALE}{\ensuremath{\mathcal{\localtext\mconnect E}}\xspace}
\newcommand{\LOCALS}{\ensuremath{\mathcal{\localtext\mconnect S}}\xspace}
\newcommand{\LOCALQ}{\ensuremath{\mathcal{\localtext\mconnect Q}}\xspace}
\newcommand{\PLOCAL}{\ensuremath{\varphi\textrm{\upshape -}\mathcal{\localtext}}\xspace}
\newcommand{\LOCC}{\ensuremath{\mathcal{LOCC}}\xspace}
\def\sV{\mathscr{V}}

\newcommand{\floor}[1]{\left\lfloor{#1}\right\rfloor}

\begin{document}
\begin{titlepage}

\title{
\textbf{What Can be Observed Locally?\\{\Large Round-based Models for Quantum
  Distributed Computing}}}

\author{
Cyril Gavoille\thanks{Supported by the ANR project ``ALADDIN'', and the
  INRIA \'equipe-project ``C\'EPAGE''.}\\
\normalsize{LaBRI - University of Bordeaux}\\[-.3ex]
\normalsize{Talence, France}
\and
Adrian Kosowski$^\ast$\\
\normalsize{LaBRI - University of Bordeaux}\\[-.3ex]
\normalsize{Talence, France}
\and
\\[-2mm]
Marcin Markiewicz\\
\normalsize{Institute of Theoretical Physics and Astrophysics}\\[-.3ex]
\normalsize{University of Gda\'nsk, Poland}
}

\date{}

\maketitle
\thispagestyle{empty}

\begin{abstract}
It is a well-known fact that, by resorting to quantum processing in addition to manipulating classical information, it is possible to reduce the time
complexity of some centralized algorithms, and also to decrease the
bit size of messages exchanged in tasks requiring communication among
several agents.

Recently, several claims have been made that certain fundamental
problems of distributed computing, including \emph{Leader Election}
and \emph{Distributed Consensus}, begin to admit feasible and
efficient solutions when the model of distributed computation is
extended so as to apply quantum processing. This has been achieved
in one of two distinct ways: (1) by initializing the system in a
quantum entangled state, and/or (2) by applying quantum communication
channels. In this paper, we explain why some of these prior claims are
misleading, in the sense that they rely on changes to the model unrelated to quantum processing. On the
positive side, we consider the aforementioned quantum extensions when
applied to Linial's well-established \LOCAL model of distributed
computing.
For both types of extensions, we put forward valid proof-of-concept examples of distributed
problems whose round complexity is in fact reduced through genuinely
quantum effects, in contexts which do not depend on the anonymity of nodes.

Finally, we show that even the quantum variants of the \LOCAL model
have non-trivial limitations, captured by a very simple (purely
probabilistic) notion which we call ``physical locality''
(\PLOCAL). While this is strictly weaker than the ``computational
locality'' of the classical \LOCAL model, it nevertheless implies that
for many distributed combinatorial optimization problems, such as
\emph{Maximal Independent Set}, the best currently known lower time
bounds cannot be broken by applying quantum processing, in any
conceivable way.
\end{abstract}


\end{titlepage}

\section{Introduction}

The introduction of computational models based on quantum computing,
starting from the works of Deutsch in the 1980's~\cite{Deutsch85}, has led to the
advent of a new branch of complexity theory. Many studies have focused
on the complexity class BQP of problems solvable on a quantum computer
in polynomial time with bounded error probability (which most famously
includes the integer factorization problem~\cite{Shor94, Shor97,
  Buhrman96}), and its relation to the classical complexity
classes. On the other hand, in an even wider time-frame, properties of
quantum-mechanical systems have proven to be of interest from the
perspective of game theory~\cite{BT08, EWL99, BH01}, information theory
\cite{IkeMike,Jae07,BS98}, and distributed systems~\cite{BT08,DP08a}. One
such major line of study concerns applying quantum effects in order to
reduce communication complexity, i.e., to decrease the number of
communication bits required to solve a specific task performed in a
system with several distributed agents. When expressed in the language
of distributed computing, such research is roughly trying to address
the question: \emph{Can quantum effects be used to enhance distributed
  computations with messages of bounded size, in settings inspired by
  the \CONGEST distributed model\footnote{See \cite{Peleg99} for an
    introduction to the \CONGEST model.}?}

The quantum variant of \CONGEST, widely studied in physics, is known
as the \LOCC model\footnote{It stands for \emph{Local Operations and
    Classical Communication}.}. It exploits the key quantum-mechanical
concept of an entangled state~(see e.g.~\cite{IkeMike,
  HHH96}). This is achieved by altering the initialization phase of
the system to allow for a starting state entangled among all the
processors, which are locally given quantum computation capabilities;
however, communication between processors is still restricted to the
exchange of classical information, only. This application of
pre-entanglement has been shown to decrease the number of
communication bits required to solve certain distributed problems with
output collected from one node, and consequently, to decrease the
number of required communication rounds when message sizes are bounded
(see e.g.~\cite{CGL99} for the first proof-of-concept example, or
\cite{Zukowski08} for a survey of related results). Many other works
on the subject have focused on characterising the physical evolution
of states attainable in the \LOCC model~\cite{Nielsen99, OMM04,
  Nest2006}, while other authors have dealt with the combinatorial
complexity of distributing the entangled state over the whole system
in the initialization phase~\cite{SKP04}. Other modifications of the
model attempt to show that a denser coding of information in
transmitted messages is possible when using quantum channels, as
compared to classical communication links (see e.g.~\cite{BCdWZ99}).

\paragraph{Related work.}

Very recently, some authors have begun to study the impact of quantum
effects on fundamental concepts of the theory of distributed
computing. An overview of this line of research is contained in the
recent survey paper by Denchev and Pandurangan~\cite{DP08a}. One especially interesting result is that the leader election problem can be solved in distributed systems with quantum links, but no pre-entanglement~\cite{TKM05,KMS08}.
Some authors have also claimed that problems related to leader election~\cite{PSK03, dHP06} and distributed
consensus~\cite{dHP06, Helm08} can be solved in distributed systems aided by quantum
pre-entanglement.

This paper constitutes an attempt to provide a consistent framework
for the aforementioned discussions on distributed computing in a
quantum setting. We point out problems with some of the
work related to distributed computing with pre-entanglement, and propose a completely different perspective
for future study. Since our intention is to keep the discussion
simple, and also to focus mainly on essential questions of
\emph{locality} in a combinatorial setting, we use as the starting
point for all our considerations the well-established \LOCAL model
a.k.a.\ Linial's Free model~\cite{Linial87, Linial92}.

\paragraph{Our contribution and outline of the paper.}

In Section~\ref{sec:def} we briefly outline the \LOCAL model and its
extensions, obtained by modifying the initialization of the system
set-up and/or adding quantum communication capabilities on the
edges. Whereas this discussion is self-contained, we also provide a
formal mathematical definition of the corresponding notions in
Appendix~\ref{app:def}.

In Section~\ref{sec:com} we compare the computational power of models based on the proposed extensions of \LOCAL. In particular, we prove that adding quantum extensions to the \LOCAL model decreases the round complexity of certain distributed problems. This is achieved through simple proof-of-concept examples.

On the other hand, in Section~\ref{sec:low} we introduce a probabilistic framework for proving lower bounds on the
distributed time complexity of computational problems in any quantum (or other
unconventional) models based on \LOCAL. This is directly applied to obtain such lower
bounds for many combinatorial optimization
problems, including Maximal Independent Set, Greedy Graph Coloring,
and problems of spanner construction. As a side effect, the simple concept of ``physical
locality'' formulated in this section, leads to the definition of a computational
model we call \PLOCAL, which appears to be of independent interest.

Finally, in Section~\ref{sec:ex} we make an attempt to clarify issues
with nearly all the related work on quantum distributed computing as
surveyed by~\cite{DP08a}. We discuss previous claims of several
authors (\cite{PSK03, dHP06, Helm08, TKM05, KMS08}) which
state that problems such as Leader Election or Distributed Consensus
benefit from the application of quantum processing. We explain why
some of the statements from~\cite{PSK03, dHP06, Helm08} should be approached with caution.

Section~\ref{sec:end} contains some concluding remarks and suggests
directions of future studies.

\section{Description of Computation Models}\label{sec:def}

In this section we briefly recall the computational properties of the
\LOCAL model, which has been the subject of intensive study in the
last 20 years, starting from the seminal works
of~\cite{Linial87,NS95a}. When considering the \LOCAL model in the
context of quantum processing, it has to be noted that simply
introducing a ``quantum computer'' as a module in each processor does
not affect the power of the model, since in \LOCAL the processors as
such are already assumed to have unbounded capabilities of local
computation.

There exist two distinct and independent approaches to extending the
\LOCAL model: by modifying the initial set-up of the system (leading
to extensions which we call \SE and \EN), and by introducing quantum
communication channels (the \QC extension). Of these three extensions,
two (\EN and \QC) rely on quantum processing and roughly correspond
to settings studied in some related work~\cite{DP08a}, whereas the
third extension (\SE) is purely computational in the classical sense,
and is introduced in this work.

The discussion which follows is intentionally informal, whereas
rigorous definitions and some further considerations are postponed to
Appendix~\ref{app:def}. The formalism in the Appendix is used in
particular for showing computational limitations of models and
pointing out errors in previous work, hence we keep it precise in a
mathematical sense and free from any implicit assumptions.

\paragraph{The \LOCAL model.}

It is assumed that the distributed system consists of a set of
processors $V$ (with $|V|=n$) and operates in a sequence of
synchronous rounds, each of which involves unbounded computations on
the local state variables of the processors, and a subsequent exchange
of messages of arbitrary size between pairs of processors which are
connected by links (except for round 0, which involves local
computations, only). Nodes can identify their neighbours using integer
labels assigned successively to communication ports. The local
computation procedures encoded in all processors are necessarily the
same, and initially all local state variables have the same value for
all processors, except for one distinguished local variable $x(v)$ of
each processor $v$ which encodes input data. The input of a problem is
defined in the form of a labeled graph $G_x$, where $G=(V,E)$ is the
system graph, while $x : V \to \N$ is an assignment of labels to
processors. The output of the algorithm is given in the form of a
vector of local variables $y : V \to \N$, and the algorithm is assumed
to terminate once all variables $y(v)$ are definitely fixed. Herein we
assume that faults do not appear on processors and links, that local
computation procedures may be randomized (with processors having
access to their own generators of random variables), and that the
input labels $x$ need not in general be unique for all processors.

In our considerations, it is convenient to assume that the set of
processors $V$ is given \emph{before} the input is defined. This is
used for convenience of notation, and does not affect neither the
model in any way, nor the anonymity of nodes in the considered problems.

\paragraph{Initialization of the system (\SE and \EN extensions).}

In the \LOCAL model, it is assumed that the initial set-up of all the
processors is identical. This assumption can be relaxed by allowing
the processors to obtain some information from a central helper, but
only before the start of the distributed process (i.e., independently
of the input $G_x$). The initialization procedure is an integral part
of the algorithm used for solving the distributed problem. Several
different forms of initialization can be naturally defined; for
clarity of discussion, we consider only two extensions of the model:
the \SE extension (for \emph{Separable} state), which allows for the
most general form of initialization possible in a classical
computational setting, and the more powerful \EN extension (for
\emph{Entangled} state), which allows for the most general form of
initialization available in a quantum distributed system.

\subparagraph{The \SE extension.}
We say that a computational model is equipped with the \emph{\SE
extension} if the following modifications are introduced:

\begin{itemize}

\item For any computational problem, the computational procedure
  consists of the distributed algorithm applied by all the processors
  during the rounds of computation, and an additional (randomized)
  procedure executed in a centralized way in the initialization phase.
  The result of the initialization procedure is an assignment $h : V
  \to \N$ of \emph{helper} variables to the set of processors. The
  helper variables are independent\footnote{Helper variables that do
  depend on the inputs are referred to in the literature as
  \emph{Oracles}~\cite{FIP06,FGIP07}. Such extensions are not
  discussed in this paper.} of the input $G_x$.

\item For each processor $v\in V$, at the start of round 0, its input
  label $x(v)$ is augmented by the value $h(v)$, stored in a helper
  register of the local memory.

\end{itemize}

It is straightforward to show that the above formulation has two
equivalent characterizations. From a computational perspective, we may
equivalently say that for each processor $v$, the helper
initialization value $h(v)$ encodes: (1) a unique identifier of $v$
from the range $\{1,\ldots,n\}$, (2) the value of $n$, (3) the value
of a random number, chosen from an arbitrarily large range, and shared
by all processors. All further helper information is unnecessary,
since it can be computed by the processors in round 0 of the
distributed computations.

Alternatively, we may say that through the randomized initialization,
according to some probability distribution we choose some
deterministic initialization of the set of states of individual
processors. This intuition precisely corresponds to the notion of a
state with uncertainty in classical statistical physics, referred to
in quantum-mechanical discussions as a (mixed) \emph{separable state}
of the system. It is obviously true to say that \emph{whenever a
problem is solved in a model with the \SE extension, it may benefit
solely from the modification of the system initialization, and not
from the laws of quantum mechanics}.

\subparagraph{The \EN extension.}
Unlike in classical physics, in quantum mechanics not every
initialization of the system has to follow the above pattern. Consider
a scenario in which we centrally create an initial global state of the
whole system of processors, and spatially distribute ``parts'' of it
to the individual processors (for example, by sharing out among the
nodes a set of quantum-correlated photons, coming from a single
SPDC\footnote{Spontaneous Parametric Down-Conversion.}  emission
process). Then, each of the processors can perform operations on the
``part'' of the state assigned to its spatial location; by a loose
analogy to processing of classical information, this is sometimes
referred to as each processor ``manipulating its own quantum
bits (qubits)''. Given a general initial state of the system,
the outcome of such a physical process, as determined by the
processors, may display correlations which cannot be described using
any classical probabilistic framework. Initial states which can be
lead to display such properties are called non-separable, or
\emph{entangled states}. Quantum entanglement is without doubt one of
the predominant topics studied in quantum-mechanical literature of the
last decades; we refer the interested reader to e.g.~\cite{IkeMike}
for an extensive introduction to the topic.

We say that a computational model is equipped with the \emph{\EN
extension} if all processors are equipped with helper quantum
information registers $h$, and the computational procedure used to
solve a problem sets in the initialization phase in a centralized way
some chosen, possibly entangled, quantum state over the set of quantum
information registers $h$ of all processors, in a way independent of
the input graph $G_x$.

Of course, the definition of the \EN extension does not require that the starting state is entangled; for the special case when it is separable, the \EN extension is precisely equivalent to the \SE extension.

\paragraph{Communication capabilities (\QC extension).}

Whereas the application of local quantum operations in each processor
does not increase the power of the \LOCAL model as such, the situation
changes when the processors can interact with each other using quantum
communication channels. Intuitively, such channels allow for the
distribution of an entangled state by a processor over several of its
neighbours in one communication round; such an effect cannot be
achieved using classical communication links.

We say that a computational model is equipped with the \emph{\QC
extension} if all communication links between processors in the system
graph are replaced by quantum communication channels.

\paragraph{Models based on extensions.}

Modifications to the initialization and communication capabilities of
the system are completely independent of each other. For
initialization, we can apply no extension, use a separable state
(\SE), or an entangled state (\EN). For communication, we can apply no
extension (message exchanges with classical information), or use
quantum channels (\QC). Hence, we obtain $6$ possible models (\LOCAL,
\LOCALS, \LOCALE, \LOCALQ, \LOCALQS, \LOCALQE), which are discussed in
the following section. Some of these collapse onto each other, in
particular, \LOCALQE and \LOCALE are equivalent in terms of
computational power (Proposition~\ref{pro:eqe}).

\section{Hierarchy of Quantum Models}\label{sec:com}

\subsection*{Comparing the Power of Computational Models}

In order to compare the computational power of different models, we
introduce two basic notions: that of the \emph{problem} being solved,
and of an \emph{outcome} of the computational process.

\begin{definition}
A \emph{problem} $\mathscr P$ is a mapping $G_x \mapsto \{y^i\}$,
which assigns to each input graph $G_x$ a set of permissable output
vectors $y^i : V \to \N$.
\end{definition}

Instead of explicitly saying that we are interested in finding
efficient (possibly randomized) distributed algorithms for solving
problems within the considered computational models, we characterize
the behavior of such procedures through the probability distribution
of output vectors which they may lead to, known as an
\emph{outcome}. In fact, such a probability distribution is
necessarily well defined, whereas formally describing the
computational process may be difficult in some unconventional settings
(see e.g.~the \PLOCAL model in Section~\ref{sec:low}).

\begin{definition}
An \emph{outcome} $\mathscr O$ is a mapping $G_x \mapsto \{(y^i,
p^i)\}$, which assigns to each input graph $G_x$ a normalized discrete
probability distribution $\{p^i\}$, such that: $\forall_i\ p^i >0$ and
$\sum_i p^i = 1$, with $p^i$ representing the probability of obtaining
$y^i : V \to \N$ as the output vector of the distributed system.
\end{definition}

\begin{definition}
For any outcome $\mathscr O$ in a computational model $\mathcal M$
which is a variant of \LOCAL, we will write $\mathscr O\in \mathcal
M[t]$ if within model $\mathcal M$ there exists a distributed
procedure which yields outcome $\mathscr O$ after at most $t$ rounds
of computation.
\end{definition}

We will say that an outcome $\mathscr O$ is a \emph{solution} to
problem $\mathscr P$ \emph{with probability} $p$ if for all $G_x$, we
have: $\sum_{\{(y^i, p^i) \in \mathscr O(G_x)\ :\ y^i \in \mathscr
P(G_x)\}}\ p_i\ \geq\ p$. When $p=1$, we will simply call $\mathscr O$
a \emph{solution} to $\mathscr P$ \emph{(with certainty)}.

By a slight abuse of notation, for a problem $\mathscr P$ we will
write $\mathscr P\in \mathcal M[t]$ (respectively, $\mathscr P\in
\mathcal M[t,p]$) if there exists an outcome $\mathscr O\in \mathcal
M[t]$ which is a solution to problem $\mathscr P$ (respectively, a
solution to problem $\mathscr P$ with probability $p$).

For two computational models $\mathcal M_1$, $\mathcal M_2$, we say
that $\mathcal M_1$ is \emph{not more powerful than} $\mathcal M_2$
(denoted $\mathcal M_1 \subseteq \mathcal M_2$) if for every problem
$\mathscr P$, for all $t\in\N$ and $p>0$, $\mathscr P \in \mathcal
M_1[t,p] \implies \mathscr P \in \mathcal M_2[t,p]$. The relation
$\subseteq$ induces a partial order of models which is naturally
extended to say that $\mathcal M_1$ and $\mathcal M_2$ are
\emph{equivalent} ($\mathcal M_1 = \mathcal M_2$), or that $\mathcal
M_1$ is \emph{less powerful than} $\mathcal M_2$ ($\mathcal M_1
\subsetneq \mathcal M_2$).

It can easily be proved that $\mathcal M_1 \subseteq \mathcal M_2$ if
and only if for every outcome $\mathscr O$, for all $t\in\N$,
$\mathscr O \in \mathcal M_1[t] \implies \mathscr O \in \mathcal
M_2[t]$. Such an outcome-based characterisation of models is
occasionally more intuitive, since it is not explicitly parameterised
by probability $p$.

In all further considerations, when proving that $\mathcal M_1
\subsetneq \mathcal M_2$, we will do so in a stronger, deterministic
sense, by showing that there exist a problem $\mathscr P$ and $t\in
\N$ such that $\mathscr P \in \mathcal M_2[t]$ and $\mathscr P\notin
\mathcal M_1[t]$.

\subsection*{Relations Between Quantum Models}

\begin{figure}[t]
\centering\includegraphics[width=0.9\textwidth]{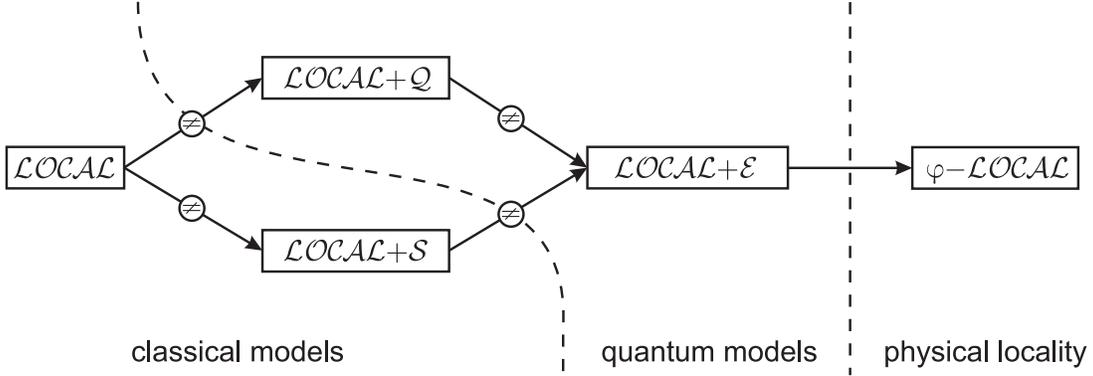}
\caption{Hierarchy of computational extensions to the \LOCAL
model. See Section~\ref{sec:low} for a definition of the \PLOCAL
model, and Section~\ref{sec:def} or Appendix~\ref{app:def} for
definitions of all other models.}\label{fig:model}
\end{figure}

The most natural variants of \LOCAL which are based on the extensions
proposed in the previous subsection are the classical model with
separable initialization (\LOCALS), and quantum models with
pre-entanglement at initialization, quantum channels, or both (\LOCALE,
\LOCALQ, and \LOCALQE, respectively). The strengths of the models can
obviously be ordered as follows: $\LOCAL \subseteq \LOCALQ \subseteq
\LOCALQS \subseteq \LOCALQE$, and $\LOCAL \subseteq \LOCALS \subseteq
\LOCALE \subseteq \LOCALQE$. We now proceed to show that, whereas
$\LOCALE = \LOCALQE$, all the remaining inclusions are in fact
strict. The hierarchy of the most important models is shown in
Fig.~\ref{fig:model}.

\begin{proposition}\label{pro:sep}
$\LOCAL \subsetneq \LOCALS$. Moreover, there exists a problem
$\mathscr P$ such that $\mathscr P\in \LOCALS[0]$ and $\mathscr
P\not\in \LOCAL[t]$ for all $t \in \mathbb{N}$.
\end{proposition}

\begin{proof}
Any problem, which can be solved when given unique node identifiers
from the range $\{1,\ldots,n\}$ is clearly in $\LOCALS[0]$. On the
other hand, there are many examples of such problems which are not in
$\LOCAL$ (or require $\Omega(n)$ rounds assuming that the system graph
is connected and node labels are unique), most trivially the problem
$\mathscr P$ of assigning unique node identifiers from the range
$\{1,\ldots,n\}$ to all nodes.
\end{proof}

More interestingly, one can show that $\LOCALS$ benefits due to the fact that helper variables $h(v)$
can encode a value which is set in a randomized way. Consider as a simple example a problem $\mathscr P'$ whose input is a graph $G=(V,E)$, of sufficiently large order $n$, with input labels of the nodes encoding unique node identifiers $\{1,\ldots,n\}$ and the value of $n$; moreover, $G$ is restricted to be the complete graph $K_{n}$ minus exactly one edge. The goal is to select an edge of the graph, i.e., output $y$ must be such that for some two nodes $u,v\in V$, with $\{u,v\} \in E$, we have $y(u)=y(v)=1$, and for all other $w\in V$ we have $y(w)=0$. Even with the knowledge of node identifiers and $n$, in the \LOCAL model the problem cannot be solved with high probability without communication, i.e., within $0$ rounds: we have $\mathscr {P'} \notin \LOCAL[0,e^{-1}]$ (the proof is technical, see Appendix~\ref{app:kne}).
On the other hand, within the \LOCALS model this problem admits a solution in 0 rounds with probability arbitrarily close to $1$ for sufficiently large $n$. Similar arguments can be applied to display the difference between the models for more advanced problems which simulate collaborative mobile agent scenarios, in particular variants of the cops-and-robbers problems in graphs.

We now point out the difference in power between the classical and quantum
models. The proofs proceed by rephrasing one of the best established
results of quantum interferometry, first introduced in the context of
the so called Bell's Theorem without inequalities, for a 3-particle
quantum entangled state (cf.~\cite{GHZ89} for the original paper,
\cite{Mermin90} for a very informal intuition, or \cite{PCZWZ08} for a
contemporary exposition). We use its more algorithmic modulo-4 sum
formulation, similar to that found in \cite{Zukowski08}.

\begin{theorem}\label{pro:ghz}
$\LOCALS \subsetneq \LOCALE$. Moreover, there exists a problem
$\mathscr P$ such that $\mathscr P\in \LOCALE[0]$ and $\mathscr
P\not\in \LOCALS[t]$ for all $t\in \mathbb{N}$.
\end{theorem}

\begin{proof}
Let $\mathscr P$ be a problem defined on a system with $3$ nodes. Let
the input graph be empty, and assume that input labels $x =
(x_1,x_2,x_3) \in \{0,1\}^3$ of respective nodes satisfy the condition
$x_1+x_2+x_3\in\{0,2\}$. An output $y = (y_1,y_2,y_3) \in \{0,1\}^3$
is considered valid for input $x$ if and only if $2(y_1 + y_2 + y_3)
\equiv (x_1 + x_2 + x_3) \mod 4$. This problem is not in \LOCALS,
since finding a solution with certainty would imply that there exist
three deterministic functions $Y_1, Y_2, Y_3 : \{0,1\} \to \{0,1\}$,
such that for any input vector $(x_1, x_2, x_3)$ satisfying the
constraints of the problem, $(Y_1(x_1), Y_2(x_2), Y_3(x_3))$ is a
valid output vector. It is immediate to show that this is impossible,
since by considering all of the possible inputs, we obtain the set of
equations: $2 (Y_1(0) + Y_2(0) + Y_3(0)) \equiv 0 \mod 4$, and $2
(Y_1(1) + Y_2(1) + Y_3(0)) \equiv 2 (Y_1(1) + Y_2(0) + Y_3(1))\equiv 2
(Y_1(0) + Y_2(1) + Y_3(1)) \equiv 2 \mod 4$, which is
contradictory: by summing the left-hand sides of all four equations we obtain $0\equiv 6 \mod 4$.


The situation is different when the system operates in the \LOCALE
model starts in an entangled state. The procedure required to obtain a
valid solution is described in detail in~\cite{GHZ89}. In brief, in the
initialization phase we share out to each of the processors one of 3
entangled qubits, carried e.g.~by photons, which are in the entangled
tripartite state known as the GHZ state (namely $\frac 1 {\sqrt2}
(|000\rangle + |111\rangle)$ in Dirac's notation for pure
states). Each of the processors then performs a simple transformation
on ``its own'' qubit, in a way dependent only on the processor's input
$x_i$. Finally, a measurement is performed, and it can be shown that
the probability distribution of obtained output vectors (the outcome)
is that stated in Table~\ref{tab:ghz}. Since all of the outputs are
accepted as valid for the considered problem $\mathscr P$, this
implies that $\mathscr P \in \LOCALE[0]$.
\begin{table}
\caption{An outcome $\mathscr O$ which is a solution (with certainty)
to the modulo-4 sum problem on the 3-node empty graph, and belongs to
$\LOCALE[0]$ (see Theorem~\ref{pro:ghz}).\newline}\label{tab:ghz}
\centering
\begin{tabular}{c|cc}
\toprule
Input & Probability & Output\\
$(x_1, x_2, x_3)$ & $p^i$ & $(y_1^i, y_2^i, y_3^i)$\\
\midrule
(0, 0, 0) & $ 1/4$ & (0, 0, 0)\\
& $ 1/4$ & (0, 1, 1)\\
& $ 1/4$ & (1, 0, 1)\\
& $ 1/4$ & (1, 1, 0)\\
\bottomrule
\end{tabular}
\hspace{5mm}
\begin{tabular}{c|cc}
\toprule
Input & Probability & Output\\
$(x_1, x_2, x_3)$ & $p^i$ & $(y_1^i, y_2^i, y_3^i)$\\
\midrule
\multirow{4}{2cm}{\phantom{or} (0, 1, 1)\newline or (1, 0, 1) \newline or (1, 1, 0)} & $ 1/4$ & (1, 1, 1)\\
& $ 1/4$ & (1, 0, 0)\\
& $ 1/4$ & (0, 1, 0)\\
& $ 1/4$ & (0, 0, 1)\\
\bottomrule
\end{tabular}
\end{table}
\end{proof}

We note that the obtained outcome $\mathscr O \in \LOCALE[0]$ is a
solution to $\mathscr P$ with certainty, but it is not deterministic,
yielding different outputs with probability $1/4$
(Table~\ref{tab:ghz}); in fact, within \LOCALE there does not exist an
outcome which is a solution to $\mathscr P$, and yields some output
with probability $1$. This sort of situation could not occur in
\LOCAL, or in any other classical model.

\begin{proposition}\label{pro:q}
$\LOCAL \subsetneq \LOCALQ$. Moreover, for any $t>0$, there exists a
problem $\mathscr P$ such that $\mathscr P\in \LOCALQ[t]$ and
$\mathscr P\not\in \LOCAL[2t-1]$.
\end{proposition}

\begin{proof}
The proof proceeds by a modification of the argument from
Theorem~\ref{pro:ghz}. This time, we consider a system on $n=3k+1$
nodes, and an input graph with the topology of a uniformly subdivided
star with a central node of degree $3$. The modified problem $\mathscr
P'$ consists in solving the problem from Theorem~\ref{pro:ghz},
when the three input and output values are put on the three leaves of
the star. Within \LOCAL, this problem requires $2k$ rounds to solve,
since the three leaves are at a distance of $2k$ from each other, and
need to communicate to solve the problem. On the other hand, in
\LOCALQ we are given quantum communication links. Hence, in round 0,
the central node can create an entangled tripartite GHZ state, and
propagate its qubits in $k$ rounds\footnote{Observe that intermediate
nodes simply send the qubit on, without making any copies.} to the leaves of the graph,
which then apply the previously discussed quantum procedure.
\end{proof}

Whereas the time distinction between \LOCALS and \LOCALE given by
Theorem~\ref{pro:ghz} is remarkable (since it considers the
feasibility of solving problems, or when discussing connected graphs,
a speed-up from $\Omega(n)$ to $0$ communication rounds), the
situation is less clear between \LOCALQ and \LOCAL. Although a
speed-up factor of~2 as expressed by Proposition~\ref{pro:q} looks
like a natural limit, the authors know of no conclusive arguments to
show that it cannot be increased further.

Finally, following the argumentation of~\cite{DP08a}, we note that
$\LOCALE = \LOCALQE$, or in other words that, given access to
pre-entanglement, it is possible to simulate quantum links by means of
classical ones. The effect used to achieve this is known as quantum
teleportation~\cite{PCZWZ08}; by carefully choosing an entangled state over
the whole system, it can be applied even when the communicating nodes
do not yet know their neighbors' unique identifiers.



\begin{proposition}[\cite{DP08a}]\label{pro:eqe}
$\LOCALE = \LOCALQE$.
\end{proposition}

To complete a discussion of Fig.~\ref{fig:model}, we point out that
$\LOCALQ$ is incomparable with $\LOCALS$. This is because the problem
discussed in the proof of Proposition~\ref{pro:sep} belongs to
$\LOCALS$, but not to $\LOCALQ$, and the problem discussed in the
proof of Proposition~\ref{pro:q} belongs to $\LOCALQ[1]$, but not to
$\LOCALS[1]$.

The $\LOCALQS$ model has been left out from discussion, since it
appears to be of little significance. By considering the same problems
as before, we have $\LOCALQS \subsetneq \LOCALQE = \LOCALE$, so
\LOCALQS could be placed directly to the left of $\LOCALE$ in
Fig.~\ref{fig:model}.

\section{Lower Time Bounds Based on Physical Locality (\PLOCAL)}\label{sec:low}


Proving lower bounds on the power of quantum models is
problematic. This results, in particular, from the fact that there
does not exist as yet an easy-to-use classification of entangled
states, or of quantum operations (completely positive maps) which can be
performed to transform one quantum state into another. However, in the
context of distributed computing, it is possible to consider a more
general framework of physical locality, leading to the \PLOCAL model
we define hereafter, which in turn can be used to bound the power of
quantum models.

Within the classical \LOCAL model, we can say that the output of any
processor $v$ after $t$ rounds has to be computed based on the input data which
can be collected from the input graph $G_x$ by performing an
exploration up to a depth of $t$, starting from node $v$; we call this
the \emph{distance-$t$ local view} denoted by $\sV_t(G_x,v)$. This leads to a simple characterisation of the \LOCAL model in terms of valid outcomes (see Appendix~\ref{app:local} for a formalization).

In order to allow for quantum extensions to local, the assumption of classical computability needs to be relaxed, while at the same time retaining in some form the assumption of locality, since it is an essential part of physical
theory as we understand it today (cf. e.g.~\cite{Shimony84, Streater07}
for different approaches to the problem). To define locality, for a
moment we choose to look at the system from a physicist's perspective,
with the distributed system as an experimental stand, with processors
as black boxes, with input data $G_x$ as part of the experimental
set-up, and with output $y$ as the data resulting of a single
experiment. For each input, the experiment is performed for an
ensemble of identical systems, obtaining a probability distribution of
outputs $\{(y^i, p^i)\}$. Now, given a round-based model with
interactions between nearest neighbors only, the physical
understanding of locality is as follows: \emph{Locality is violated if
  and only if, based on the available output data, we can conclusively
  verify that after $t$ rounds some subset $S$ of processors was
  affected by input data initially localized outside its view
  $\sV_t(G_x, S) := \bigcup_{v \in S} \sV_t(G_x,v)$.}

Using the above intuition, we now formalize this notion to obtain what
we call the \PLOCAL model, i.e., the weakest possible distributed
model which still preserves physical locality. Given an output
distribution $\{(y^i, p^i)\}$ acting on $V$, for any subset of
vertices $S\subseteq V$ we define its \emph{marginal distribution on
set} $S$, $\{(y^i, p^i)\}[S]$, as the unique distribution
$\{(\overline y^j, \overline p^j)\}$ acting on $S$ which satisfies the
condition $\overline p^j = \sum_{\{i\ :\ \overline y^j = y^i[S]\}}
p^i$, where $y^i[S]$ is the restriction of output $y^i : V \to \N$ to nodes from subset $S\subseteq V$.

\begin{definition}
An outcome $G_x\mapsto\{(y^i, p^i)\}$ belongs to $\PLOCAL[t]$
if for all subsets $S\subseteq V$, for any pair of inputs $G_x^{(a)}$,
$G_x^{(b)}$ such that $\sV_t(G_x^{(a)},S) = \sV_t(G_x^{(b)},S)$, the
output distributions corresponding to these inputs have identical
marginal distributions on set $S$, i.e., $\{(y^{i(a)}, p^{i(a)})\}[S]
= \{(y^{i(b)}, p^{i(b)})\}[S]$.
\end{definition}

Quantum relaxations of the \LOCAL model, whether obtained through
application of pre-entanglement, quantum channels, or both, lie in
terms of strength ``in between'' the \LOCAL and \PLOCAL model. This is
expressed by the following theorem, whose proof we defer to
Appendix~\ref{app:plocal}.

\begin{theorem}
$\LOCALQE \subseteq \PLOCAL$.
\end{theorem}

The theorem captures the property of locality of nearest-neighbor
interactions in quantum mechanics, and its proof can be seen as a
boundary case (for discrete rounds) of the more physical
continuous-time setting studied in~\cite{RobBra78:2}. It does not
rely in any way on any other physical concepts, such as causality or
speed of information in the theory of relativity.

Although it is not clear whether the containment in the above theorem
is strict (we leave this as an open question), the \PLOCAL model is
still sufficiently constrained to preserve many important lower time
bounds known from the \LOCAL model, which are based on arguments of
indistinguishability of local views of a node for different inputs. In
particular, by careful analysis, it is easy to prove the following
statements for the \PLOCAL model.

\begin{itemize}
    \item The problem of finding a maximal independent set in the
    system graph requires $\Omega(\sqrt \frac{\log n }{\log \log n})$
    rounds to solve~\cite{KMW04}.
    \item The problem of finding a locally minimal (greedy) coloring
    of the system graph requires $\Omega( \frac{\log n }{\log \log
    n})$ rounds to solve~\cite{GKKN07,GKKKN09}.
    \item The problem of finding a connected subgraph with
      $O(n^{1+1/k})$ edges requires $\Omega(k)$ rounds to
      solve~\cite{DGPV08b,Elkin07}.
\end{itemize}

The matter is less clear in the case of the $(\Delta+1)$-coloring
problem. The proof of the famous lower bound of $\frac{1}{2} \log^*{n}
- O(1)$ rounds~\cite{Linial92} (and its extension to randomized
algorithms~\cite{Naor91}) does not appear to generalize from the
\LOCAL model to the \PLOCAL model; we are unaware of any (even
constant) bound on the number of rounds required to find a solution to
$(\Delta+1)$-coloring in \PLOCAL. Some indication that the technique
of coloring neighborhood graphs, used by Linial, may not apply in
\PLOCAL, is that this technique can likewise be used to show a lower
bound of $\floor{\frac{n}{2}} - 1$ rounds on the time required for
$2$-coloring the cycle $C_{n}$, where $n$ is even. However, in \PLOCAL
the same problem admits a solution in fewer rounds.

\begin{theorem}
The problem of $2$-coloring the even cycle $C_n$ (given unique node
labels $x$) belongs to $\PLOCAL[\lceil \frac{n-2}{4}\rceil]$, but does
not belong to $\PLOCAL[\lceil \frac{n-2}{4}\rceil-1]$.
\end{theorem}

\begin{proof}[Proof (sketch)]
For the lower bound, consider the local view of two nodes $u,v$ which
still have disjoint views after $\lceil \frac{n-2}{4}\rceil-1$
rounds. There are at least two nodes which belong to neither the view
of $u$ nor the view of $v$; hence, $u$ and $v$ cannot distinguish
whether they are at an even or at an odd distance from each other in
the cycle. This directly leads to the lower bound, since the
definition condition of \PLOCAL can be shown to be violated for $S=\{u,v\}$.

The upper bound is generated by on outcome $\mathscr O$ of the
$2$-coloring problem, given as follows: each of the $2$ legal
$2$-colorings of $C_n$ is used as the output with probability $\frac 1
2$. Such an outcome $\mathscr O$ belongs to $\PLOCAL[\lceil
\frac{n-2}{4}\rceil]$. This can be easily verified, since for any
subset $S \subseteq V$ we either have that $S$ consists of exactly two
antipodal nodes of $C_n$, or the view $\sV_{\lceil
\frac{n-2}{4}\rceil}(C_{n_x},S)$ is simply an arc of the cycle.
\end{proof}

It would be interesting to find a constructive quantum procedure for
finding a $2$-coloring of $C_n$ in $\lceil \frac{n-2}{4}\rceil$
rounds. In particular, we have that $2$-coloring of $C_6$ belongs to
$\PLOCAL[1]$, does not belong to $\LOCALS[1]$, and do not know if it
belongs to $\LOCALE[1]$.

\section{Simple Problems in a Quantum Setting}\label{sec:ex}

In this section, we have a look at some of the related work
on quantum distributed problems, as outlined in the
survey~\cite{DP08a}. Whereas the discussion in this section relies on
the results and notation from the preceding sections, it can also be
translated into the (not always precisely described) computational
models studied in the considered related work.

Two problems which have been used to exhibit the difference between
quantum models and non-quantum models are \algo{Leader Election},
where the goal is for exactly one node of the system graph to output a
value of 1 whereas all other nodes output 0, and a problem which we
will call \algo{Bit Picking}, where the goal is for all nodes to
return the same output value, either 0 or 1.\footnote{Two sets of
  authors~\cite{dHP06, Helm08} confuse the latter question with that
  of the \algo{Distributed Consensus} problem~\cite{Lynch97, AW04}.}
These discussions include the concept of \emph{fairness}, which in the
terminology of this paper means that we are asking not about the
problems as such, but about obtaining specific (fair)
\emph{outcomes}. More precisely, we will say that \algo{Fair Leader
  Election} is the outcome which puts a uniform probability
distribution on the $n$ distinct outputs valid for \algo{Leader
  Election} (i.e., on all possible leaders), and \algo{Fair Bit
  Picking} is the outcome which puts a uniform probability
distribution on the $2$ distinct outputs valid for \algo{Bit Picking}
(i.e., picking 0 or 1).

The focus of~\cite{PSK03, dHP06, Helm08} is to show that \algo{Fair
  Bit Picking} and \algo{Fair Leader Election} belong to $\LOCALE[0]$
(even with some additional restrictions on the amount of allowed
pre-entanglement), whereas they do not belong to $\LOCAL[0]$. This
statement is correct, however, this effect is due to \emph{the
  modification of initialization of the system, and not to quantum
  mechanics}. In fact, we can make the following obvious statement.

\begin{proposition}
\algo{Fair Bit Picking} and \algo{Fair Leader Election} belong to the
non-quantum class $\LOCALS[0]$. Moreover, they can be solved with only
one bit of helper information per node, at initialization.
\end{proposition}

\begin{proof}
There is no input for the considered outcomes, hence the
initialization procedure can be defined so as to encode the
appropriate output vector in the helper data $h(v)$, choosing specific
outputs according to the required probability distribution.
\end{proof}

Consequently, this sort of study should be considered in the context
of the \LOCALS modification, or in other words, the benefits of adding
purely classical helper information to the \LOCAL model
(and not what some authors refer to as ``quantum
non-locality''). Whereas in a formal sense it is not a mistake to say
that such an effect can also be obtained when using a quantum entangled
state as the ``helper'', this is technologically difficult to implement, complicates
the discussion, and does not save information in any way, since for
the considered outcomes, the required helper data can already be encoded
using one bit per node within \LOCALS. As such, this sort of approach can be seen
as useless from the perspective of distributed
computing.\footnote{Some observations can perhaps be of purely
  physical interest. For example, for a problem with empty input,
  whenever there exists a one-bit helper function $h:V\to \{0,1\}$
  leading to the desired output in $\LOCALS[0]$, unique up to negation
  of $0$ and $1$, there will also exist an entangled pure $n$-qubit
  state spread over the nodes leading to the desired output in
  $\LOCALE[0]$, unique up to transformation of the local basis
  $\{|0\rangle,|1\rangle\}$. \cite{dHP06} note that this is the case
  for \algo{Fair Bit Picking} and \algo{Fair Leader Election}.} In
order to capture the benefit coming from the quantum
setup, one has to display quantum correlations which cannot
be described in the classical framework (Theorem~\ref{pro:ghz}
and Proposition~\ref{pro:q}).

As a side note, we mention that a separate question concerns the anonymity of nodes in the
system. Whereas it is impossible to solve leader election with
certainty in the classical anonymous setting, \cite{TKM05,KMS08}
have considered leader election in the anonymous quantum setting of \LOCALQ,
providing a nice and efficient algorithm, which obtains a valid solution with certainty.

\begin{theorem}[\cite{TKM05,KMS08}]
\algo{Leader Election} $\in \LOCALQ[n]$, for anonymous nodes.
\end{theorem}

Finally, we relate to the recent claims that the \algo{Distributed
  Consensus} can be solved in a quantum setting without
communication. Whereas these claims result from a misunderstanding of
the definition~\cite{Lynch97, AW04} of \algo{Distributed Consensus},
we point out that such a result is impossible in any quantum model,
since it is even impossible in \PLOCAL. We recall that in
\algo{Distributed Consensus}, given an assignment of input labels
$(x_1,\ldots,x_n)$ to particular processors, the goal is to obtain an
output vector $(y,\ldots,y)$, such that $y\in \{x_1,\ldots,x_n\}$.

\begin{proposition}
\algo{Distributed Consensus} $\notin \PLOCAL[0]$.
\end{proposition}

\begin{proof}
Consider a system with only two processors, having inputs $x_1, x_2
\in \{0,1\}$. Let outcome $\mathscr O$ be a valid solution to
\algo{Distributed Consensus} with certainty. Then, $\mathscr O$ must
be given as the following mapping $x \mapsto \{(p^i, y^i)\}$ for some
probability values $p,q \in [0,1]$: $(0,0) \mapsto \{(1, (0,0))\}$,
$(1,1) \mapsto \{(1, (1,1))\}$, $(0,1) \mapsto \{(p, (0,0)), (1-p,
(1,1))\}$, and $(1,0) \mapsto \{(q, (0,0)), (1-q, (1,1))\}$. Now,
suppose that $\mathscr O \in \PLOCAL[0]$. Applying the definition of
\PLOCAL to set $S$ consisting of processor $1$ only, considering
inputs $x_a = (0,0)$ and $x_b = (0,1)$, we obtain $p=1$. Likewise,
applying the same definition to set $S'$ consisting of processor $2$
only, considering inputs $x'_a = (1,1)$ and $x'_b = (0,1)$, we obtain
$p=0$, a contradiction.
\end{proof}

\section{Conclusions and Future Work}\label{sec:end}

We have pointed out that the computational power of quantum variants
of the \LOCAL model is \emph{strictly greater} than that of the
classical \LOCAL model, or that of the \LOCAL model equipped with
helper information such as a pool of shared random bits. It remains to
be seen whether a difference can be observed for any problems of
practical significance. It is potentially possible
that certain combinatorial optimization problems may benefit from
quantum extensions to the \LOCAL model. However, we can say
that the ``view-based'' limitations of the \LOCAL model still hold in
quantum models. So, one specific question which remains open is
whether the $(\Delta+1)$-Coloring problem can be solved in a
constant number of rounds in any of the relaxed variants of \LOCAL.

Finally, we can ask about a characterization of the limitations of
quantum computability, the most natural question being to establish
whether the containment $\LOCALE \subseteq \PLOCAL$ is strict. As a
matter of fact, further studies of the \PLOCAL model, which can be
seen as the weakest distributed local model, capturing verifiability
rather than computability of outcomes, appear to be of interest in
their own right.

\paragraph{Acknowledgment:}
\enlargethispage{\baselineskip}

We gratefully thank Pierre Fraigniaud and Zvi Lotker for their
preliminary discussions on the EPR effect and its applicability to
Distributed Computing. We thank Robert Alicki and W\l{}adys\l{}aw Adam Majewski for helpful discussions
concerning quantum dynamic maps, and Marek \.Zukowski for providing us with
several references on quantum information.

\newpage
\pagenumbering{roman}

\bibliographystyle{alphabbrv}

{\small
\bibliography{bib}

\newcommand{\etalchar}[1]{$^{#1}$}
\begin{thebibliography}{BCdWZ99}

\bibitem[AW04]{AW04}
H.~Attiya and J.~Welch.
\newblock {\em Distributed Computing: Fundamentals, Simulations, and Advanced
  Topics (2nd ed.)}.
\newblock Wiley-Interscience Publication, 2004.

\bibitem[Bae87]{Baez87}
J.~Baez.
\newblock Bell's inequality for {C*}-algebras.
\newblock {\em Letters in Mathematical Physics}, 13(2):136--137, February 1987.

\bibitem[BCdWZ99]{BCdWZ99}
H.~Buhrman, R.~Cleve, R.~de~Wolf, and C.~Zalka.
\newblock Bounds for small-error and zero-error quantum algorithms.
\newblock In {\em $40^{th}$ Annual IEEE Symposium on Foundations of Computer
  Science (FOCS)}, pages 358--368. IEEE Computer Society Press, October 1999.

\bibitem[BH01]{BH01}
S.~C. Benjamin and P.~M. Hayden.
\newblock Multiplayer quantum games.
\newblock {\em Physical Review A}, 64(3):030301, 2001.

\bibitem[BR79]{RobBra78}
O.~Bratelli and D.~W. Robinson.
\newblock {\em Operator algebras and quantum statistical mechanics}, volume~I.
\newblock Springer-Verlag, 1979.

\bibitem[BR81]{RobBra78:2}
O.~Bratelli and D.~W. Robinson.
\newblock {\em Operator algebras and quantum statistical mechanics}, volume~II.
\newblock Springer-Verlag, 1981.

\bibitem[BS98]{BS98}
C.~H. Bennett and P.~W. Shor.
\newblock Quantum information theory.
\newblock {\em IEEE Transactions on Information Theory}, 44:2724--2742, 1998.

\bibitem[BT08]{BT08}
A.~Broadbent and A.~Tapp.
\newblock Can quantum mechanics help distributed computing?
\newblock {\em ACM SIGACT News - Distributed Computing Column}, 39(3):67--76,
  September 2008.

\bibitem[Buh96]{Buhrman96}
H.~Buhrman.
\newblock A short note on {S}hor's factoring algorithm.
\newblock {\em ACM SIGACT News}, 27(1):89--90, March 1996.

\bibitem[CGL99]{CGL99}
R.~Cleve, D.~Gottesman, and H.-K. Lo.
\newblock How to share a quantum secret.
\newblock {\em Physical Review Letters}, 83(3):648--651, July 1999.

\bibitem[Deu85]{Deutsch85}
D.~Deutsch.
\newblock Quantum theory, the {C}hurch-{T}uring principle and the universal
  quantum computer.
\newblock {\em Proceedings of the Royal Society of London}, A400:97--117, 1985.

\bibitem[DGPV08]{DGPV08b}
B.~Derbel, C.~Gavoille, D.~Peleg, and L.~Viennot.
\newblock On the locality of distributed sparse spanner construction.
\newblock In {\em $27^{th}$ Annual ACM Symposium on Principles of Distributed
  Computing (PODC)}, pages 273--282. ACM Press, August 2008.

\bibitem[dNDVB07]{Nest2006}
M.~V. den Nest, W.~D\"ur, G.~Vidal, and H.~Briegel.
\newblock Classical simulation versus universality in measurement-based quantum
  computation.
\newblock {\em Physical Review A}, 75(1):012337, 2007.

\bibitem[DP06]{dHP06}
E.~D'Hondt and P.~Panangaden.
\newblock The computational power of the {W} and {GHZ} states.
\newblock {\em Quantum Information and Computation}, 6(2):173--183, March 2006.

\bibitem[DP08]{DP08a}
V.~S. Denchev and G.~Pandurangan.
\newblock Distributed quantum computing: A new frontier in distributed systems
  or science fiction?
\newblock {\em ACM SIGACT News - Distributed Computing Column}, 39(3):77--95,
  September 2008.

\bibitem[Elk07]{Elkin07}
M.~Elkin.
\newblock A near-optimal fully dynamic distributed algorithm for maintaining
  sparse spanners.
\newblock In {\em $26^{th}$ Annual ACM Symposium on Principles of Distributed
  Computing (PODC)}, pages 195--204. ACM Press, August 2007.

\bibitem[EWL99]{EWL99}
J.~Eisert, M.~Wilkens, and M.~Lewenstein.
\newblock Quantum games and quantum strategies.
\newblock {\em Physical Review Letters}, 83(11):3077, 1999.

\bibitem[FGIP07]{FGIP07}
P.~Fraigniaud, C.~Gavoille, D.~Ilcinkas, and A.~Pelc.
\newblock Distributed computing with advice: Information sensitivity of graph
  coloring.
\newblock In {\em $34^{th}$ International Colloquium on Automata, Languages and
  Programming (ICALP)}, volume 4596 of Lecture Notes in Computer Science, pages
  231--242. Springer, July 2007.

\bibitem[FIP06]{FIP06}
P.~Fraigniaud, D.~Ilcinkas, and A.~Pelc.
\newblock Oracle size: a new measure of difficulty for communication tasks.
\newblock In {\em $25^{th}$ Annual ACM Symposium on Principles of Distributed
  Computing (PODC)}, pages 179--187. ACM Press, July 2006.

\bibitem[GHZ89]{GHZ89}
D.~M. Greenberger, M.~A. Horne, and A.~Zeilinger.
\newblock Going beyond {B}ell's {T}heorem.
\newblock In {\em Bell's Theorem, Quantum Theory, and Conceptions of the
  Universe}, pages 69--72. Kluwer, 1989.

\bibitem[GKK{\etalchar{+}}09]{GKKKN09}
C.~Gavoille, R.~Klasing, A.~Kosowski, {\L}.~Kuszner, and A.~Navarra.
\newblock On the complexity of distributed graph coloring with local minimality
  constraints.
\newblock {\em Networks}, 2009.
\newblock To appear.

\bibitem[GKKN07]{GKKN07}
C.~Gavoille, R.~Klasing, A.~Kosowski, and A.~Navarra.
\newblock Brief announcement: On the complexity of distributed greedy coloring.
\newblock In {\em $21^{st}$ International Symposium on Distributed Computing
  (DISC)}, volume 4731 of Lecture Notes in Computer Science, pages 482--484.
  Springer, September 2007.

\bibitem[Hel08]{Helm08}
L.~Helm.
\newblock Brief announcement: {Q}uantum distributed consensus.
\newblock In {\em $27^{th}$ Annual ACM Symposium on Principles of Distributed
  Computing (PODC)}, pages 445--445. ACM Press, August 2008.

\bibitem[HHH96]{HHH96}
M.~Horodecki, P.~Horodecki, and R.~Horodecki.
\newblock Separability of mixed states: necessary and sufficient conditions.
\newblock {\em Physics Letters A}, 223:1--8, 1996.

\bibitem[Jae07]{Jae07}
G.~Jaeger.
\newblock {\em Quantum Information. An Overview}.
\newblock Springer-Verlag, 2007.

\bibitem[KMT08]{KMS08}
H.~Kobayashi, K.~Matsumoto, and S.~Tani.
\newblock Fast exact quantum leader election on anonymous rings.
\newblock In {\em $8^{th}$ Asian Conference on Quantum Information Science
  (AQIS)}, pages 157--158, August 2008.

\bibitem[KMW04]{KMW04}
F.~Kuhn, T.~Moscibroda, and R.~Wattenhofer.
\newblock What cannot be computed locally!
\newblock In {\em $23^{rd}$ Annual ACM Symposium on Principles of Distributed
  Computing (PODC)}, pages 300--309. ACM Press, July 2004.

\bibitem[Lin87]{Linial87}
N.~Linial.
\newblock Distributive graph algorithms - {G}lobal solutions from local data.
\newblock In {\em $28^{th}$ Annual IEEE Symposium on Foundations of Computer
  Science (FOCS)}, pages 331--335. IEEE Computer Society Press, October 1987.

\bibitem[Lin92]{Linial92}
N.~Linial.
\newblock Locality in distributed graphs algorithms.
\newblock {\em SIAM Journal on Computing}, 21(1):193--201, 1992.

\bibitem[Lyn97]{Lynch97}
N.~Lynch.
\newblock {\em Distributed Algorithms}.
\newblock Morgan Kaufmann Publishers, 1997.

\bibitem[Mer90]{Mermin90}
N.~D. Mermin.
\newblock Quantum mysteries revisited.
\newblock {\em American Journal of Physics}, 58(8):731--734, 1990.

\bibitem[Nao91]{Naor91}
M.~Naor.
\newblock A lower bound on probabilistic algorithms for distributive ring
  coloring.
\newblock {\em SIAM Journal on Discrete Mathematics}, 4(3):409--412, 1991.

\bibitem[NC00]{IkeMike}
M.~Nielsen and I.~Chuang.
\newblock {\em Quantum Computation and Quantum Information}.
\newblock {C}ambridge {U}niversity {P}ress, Cambridge, 2000.

\bibitem[Nie99]{Nielsen99}
M.~Nielsen.
\newblock Conditions for a class of entanglement transformations.
\newblock {\em Physical Review Letters}, 83(2):436--439, July 1999.

\bibitem[NS95]{NS95a}
M.~Naor and L.~Stockmeyer.
\newblock What can be computed locally.
\newblock {\em SIAM Journal on Computing}, 24(6):1259--1277, 1995.

\bibitem[OMM04]{OMM04}
M.~Owari, K.~Matsumoto, and M.~Murao.
\newblock Entanglement convertibility for infinite-dimensional pure bipartite
  states.
\newblock {\em Physical Review A}, 70(5):1--4, 2004.

\bibitem[PCZ{\etalchar{+}}08]{PCZWZ08}
J.-W. Pan, Z.-B. Chen, M.~\.{Z}ukowski, H.~Weinfurter, and A.~Zeilinger.
\newblock Multi-photon entanglement and interferometry.
\newblock Technical report, arXiv: quant-ph/0805.2853v1, May 2008.

\bibitem[Pel99]{Peleg99}
D.~Peleg.
\newblock Proximity-preserving labeling schemes and their applications.
\newblock In P.~Widmayer, G.~Neyer, and S.~Eidenbenz, editors, {\em $25^{th}$
  International Workshop on Graph-Theoretic Concepts in Computer Science (WG)},
  volume 1665 of Lecture Notes in Computer Science, pages 30--41. Springer,
  June 1999.

\bibitem[PSK03]{PSK03}
S.~P. Pal, S.~K. Singh, and S.~Kumar.
\newblock Multi-partite quantum entanglement versus randomization: Fair and
  unbiased leader election in networks.
\newblock Technical report, arXiv: quant-ph/0306195v1, June 2003.

\bibitem[Shi84]{Shimony84}
A.~Shimony.
\newblock Controllable and uncontrollable nonlocality.
\newblock In {\em International Symposium on the Foundations of Quantum
  Mechanics}, pages 130--139. The Physical Society of Japan, 1984.

\bibitem[Sho94]{Shor94}
P.~W. Shor.
\newblock Algorithms for quantum computation: Discrete log and factoring.
\newblock In {\em $35^{th}$ Annual IEEE Symposium on Foundations of Computer
  Science (FOCS)}, pages 124--134. IEEE Computer Society Press, November 1994.

\bibitem[Sho97]{Shor97}
P.~W. Shor.
\newblock Polynomial-time algorithms for prime factorization and discrete
  logarithms on a quantum computer.
\newblock {\em SIAM Journal on Computing}, 26(5):1484--1509, 1997.

\bibitem[SKP04]{SKP04}
S.~K. Singh, S.~Kumar, and S.~P. Pal.
\newblock Characterizing the combinatorics of distributed {EPR} pairs for
  multi-partite entanglement.
\newblock Technical report, arXiv: quant-ph/0306049v2, January 2004.

\bibitem[Str07]{Streater07}
R.~F. Streater.
\newblock {\em Lost Causes in and Beyond Physics}.
\newblock Springer-Verlag, 2007.

\bibitem[TKM05]{TKM05}
S.~Tani, H.~Kobayashi, and K.~Matsumoto.
\newblock Exact quantum algorithms for the leader election problem.
\newblock In {\em $22^{nd}$ Annual Symposium on Theoretical Aspects of Computer
  Science (STACS)}, volume 3404 of Lecture Notes in Computer Science, pages
  581--592. Springer, February 2005.

\bibitem[\.{Z}08]{Zukowski08}
M.~\.{Z}ukowski.
\newblock On {B}ell's {T}heorem, quantum communication, and entanglement
  detection.
\newblock In {\em Foundations of Probability and Physics 5}, August 2008.

\end{thebibliography}
}

\newpage
\appendix

\section{Model of a Quantum Distributed System}\label{app:def}

Any quantum-mechanical discussion relies on two fundamental concepts:
\emph{states} and \emph{observables}. Intuitively, a state can be
treated as a measure of knowledge about a physical system (usually
associated with some observer), whereas the set of observables encodes
the measurable properties of the system. There is a duality between
these two concepts: we can also say that the state of the system is
uniquely described through the distribution of outcomes of
measurements on all possible observables related to it. Whereas in
quantum-informational papers it is often convenient to focus on
states, we choose to adopt the approach more usual in mathematical
physics, which focuses on operator algebras (observables are operators
satisfying certain mathematical conditions). The algebraic approach
used as the basis of this model is generally accepted as the most
mathematically robust theory, and moreover it naturally encodes
concepts of quantum-mechanical locality, since operator algebras are
spatially localised (restricted to each processor), unlike the state
which is a global property of the system of all processors.

Most of the considerations in this paper can be without much loss of
generality viewed as finite-dimensional. Then, a \emph{(unital)
  $C^*$-algebra} can be introduced simply as some set of $m \times m$
matrices over complex numbers, which contains the identity matrix, and
is closed with respect to the operations of matrix multiplication,
matrix conjugation, and linear combination. Further on we rely only on
a few basic concepts which can be understood in accordance with their
standard definition for matrices, such as matrix multiplication, the
tensor product $\otimes$, and spectral decomposition. The interested
reader is referred to~\cite{RobBra78} for an explanation of more
advanced concepts related to $C^*$-algebras.

\comment{, and in particular an exposition of the
  Gelfand-Naimark-Segal construction (which provides the key link with
  the Hilbert-space based approach, by defining an embedding of a
  $C^*$-algebra in the algebra $\mathcal B(H)$ of bounded operators
  over a Hilbert space)}

\subsection{Specification of the Physical System}

The distributed set-up is given by a recipe which is well defined for
each $n\in \mathbb{N}$. For simplicity of the description, we assume
that all the processors know a value $D$, which is some arbitrarily
weak upper bound on the number of neighbours of the node in the
system, and that we consider problems for which the input and output
values are integers also bounded by $D$, i.e., $x,y : V \to
\{0,\ldots,D\}$.
\begin{itemize}
\item Each processor $v$ is described by its own copy $\mathcal A_v$ of a $C^*$-algebra $\mathcal A$, localised in an area associated with the processor.
\item The algebra $\mathcal A$ describing a processor is a complex system composed of the processor's several modules, given in the form of the following tensor product of $C^*$-algebras: $\mathcal A_v = \mathcal Q \otimes \mathcal{IO}_1 \otimes \cdots \otimes \mathcal{IO}_D$, where:
\begin{itemize}
    \item $\mathcal Q$ is a non-commutative (quantum) algebra encoding the computational characteristics (``hardware'') of the processor,
    \item $\mathcal {IO}_1, \ldots, \mathcal{IO}_D$ are isomorphic copies of an algebra $\mathcal {IO}$ representing a single input/output communication port of the processor (the algebra $\mathcal {IO}$ is commutative if and only if the channel is classical, and non-commutative if and only if the channel is a quantum one).
\end{itemize}

\item The multi-processor environment as a whole is described by the tensor product algebra of the algebras of specific processors, $\mathcal{A}^\otimes = \mathcal{A}_1\otimes \cdots \otimes \mathcal{A}_n$.

\item The state of the multi-processor system is a positive normalized linear functional acting on algebra $\mathcal A^\otimes$, of the form $\omega : \mathcal A^\otimes \to \mathbb R_+$. The state $\omega$ is defined at the time of the initial set-up of the distributed system, and can be used to encode pre-entanglement.

\item The \emph{input data} of processors, given by way of a function $x : V \to \{0,\ldots,D\}$, is fixed and for simplicity assumed to be \emph{outside} the quantum system. The \emph{algorithm} is defined by way of a family  of quantum operations (completely positive maps), $\varphi^x : A \to A$, for $0\leq x \leq D$,  which encode the local operation of a processor having $x$ as its input value.
\item The evolution (dynamics) of the system is given through a sequence of discrete rounds. The $t$-th round is subdivided into a phase in which some local transformations $\varphi_{v}$ are applied within each processor (computation phase), and a phase used for exchanging messages $\psi_e$ along edges $e$ between adjacent processors (communication phase). For an observable $A^\otimes$, initially we put $A^\otimes_{(0)} = \varphi_{1} \cdots \varphi_{n} A^\otimes$, and for all subsequent rounds, $A^\otimes_{(t+1)} = \varphi_{1} \cdots \varphi_{n} \psi_1 \cdots \psi_m A^\otimes_{(t)}$. The specific maps $\varphi_{v}$ and $\psi_e$  are formally defined as follows.
\begin{itemize}
\item Dynamic maps $\varphi_v : \mathcal A^\otimes \to \mathcal
  A^\otimes$ describe local operations performed during a round $t$ at
  each node. The map $\varphi_v$ acts only on the local algebra of
  processor $v$ depending on its input label $x(v)$, and is given by
  extension to the tensor product of the following transformation:
  $\varphi_{v}(A_1 \otimes \cdots \otimes A_v \otimes \cdots \otimes
  A_n) = A_1 \otimes \cdots \otimes \varphi^{x(v)}(A_v) \otimes \cdots
  \otimes A_n$. In each round $t$, all maps $\varphi_{v}$
  corresponding to different processors act on independent algebras
  and clearly commute (i.e., they can be executed simultaneously or reordered without changing the result).
\item Dynamic maps $\psi_e :  \mathcal A^\otimes \to \mathcal
  A^\otimes$ describe communication along edges $e$ of the system
  graph, and at the same time define the system graph. All the maps
  are induced by extension to the tensor product of the same
  transmission function for a pair of input/output ports $\psi :
  \mathcal{IO} \times \mathcal{IO} \to \mathcal{IO} \times
  \mathcal{IO}$, given simply as the exchange operation
  $\psi(X,Y)=(Y,X)$ if edge $e$ exists in the graph, and the identity
  operation $\psi(X,Y)=(X,Y)$ otherwise (signifying lack of
  communication). The map $\psi_e$ acts only on the the algebras
  corresponding to copies of $\mathcal{IO}$ for the input/output ports
  of the processors communicating along edge $e$, leaving all other
  algebras unchanged. The maps $\psi_e$ for different edges act on
  independent algebras and clearly commute (i.e., they can be executed simultaneously or reordered without changing the result).
\end{itemize}

\item The local algebra $\mathcal A$ of a processor contains one distinguished element: observable $M \in \mathcal A$, which is used for purposes of measurement. For simplicity we assume that $M$ has a discrete spectral decomposition of the form $M = \sum_{i=1}^{k} \lambda^i P^i$, with eigenvalues $\lambda^i \in \mathbb{R_+}$ and projectors $P^i \in \mathcal A$. The observable $M^\otimes \in \mathcal A^\otimes$ is given through the tensor products of particular processors' copies of observable $M$, as $M^\otimes = M \otimes \cdots\otimes M$.

\item The evolution of the system is assumed to terminate after $T$ rounds. After $T$ rounds, a standard von Neumann measurement process is applied to observable $M^\otimes_{(T)} = [M \otimes \cdots\otimes M]_{(T)}$ (the evolved observable $M^\otimes$ after $T$ rounds). This can be described as follows: by iterating over all possible tuples of values $(i_1,\ldots,i_n) \in \{1,\ldots,k\}^n$ for all $v$, with probability $p_{(i_1,\ldots,i_n)} = \omega ([P^{i_1} \otimes \cdots \otimes P^{i_n}]_{(T)} )$, the values $(\lambda^{i_1}, \ldots, \lambda^{i_n})$ are selected as the result of measurement for the respective processors $(1,\ldots,n)$. The probabilities $p_{(i_1,\ldots,i_n)}$ are understood here in the classical sense, and normalized in sum to $1$.

\item The output of the algorithm is obtained by applying a function $f : \mathbb{R_+} \to \{0,\ldots,D\}$ to the measurement results. Thus, values $(f(\lambda^{i_1}), \ldots, f(\lambda^{i_n}))$ are returned by processors $(1,\ldots,n)$, respectively.
\end{itemize}

Such a set-up is chosen for its simplicity, but obviously, there exist
several other definitions which lead to equivalent models. For
example, the input can be represented by enlarging the algebras
$\mathcal Q$ and introducing an additional factored part of the input
state. Also, for improved clarity of the model, we prefer to consider
dynamic maps, deferring other quantum operations (such as
measurements) to the final stage of the algorithm. This can be
achieved without loss of generality by sufficiently enlarging the
local algebras of the processors (see e.g.~\cite{Streater07} for a
high-level exposition). When defining algorithms in practice it may of
course be convenient to apply measurements in intermediate steps so as
to simplify formulation.

In what follows, we introduce some standard notation. A state $\omega$
is called \emph{pure} if it is extremal with respect to convex
combination of states (i.e., if $\omega = \alpha \omega_1 + (1-\alpha)
\omega_2$ for some states $\omega_1 \neq \omega_2$ and $0\leq\alpha <
1$, then $\alpha=0$). A pure state $\omega$ is said to be a
\emph{product state} over $\mathcal A_a \otimes \mathcal A_b$ if for
any $A \in \mathcal A_a$, $B \in \mathcal A_b$ we have $\omega
(A\otimes B) = \omega(A\otimes \mathbf 1) \omega(\mathbf 1 \otimes B)
\equiv \omega_a(A) \omega_b(B)$; for compactness, we simply write
$\omega = \omega_a \omega_b$.

\subsection{Details of the Computational Model}

From a computational perspective, the set-up of the system described in the previous section can be summarized as follows:
\begin{itemize}
\item The \emph{distributed algorithm} $(\omega, \{\varphi^x\}, \mathcal Q, \mathcal{IO}, M, f)$ is defined by setting the initial state $\omega$ of the system, the maps $\varphi^x$ which shape the local computations in each round, and the observable $M$ and function $f$ responsible for extracting the output from the quantum system. The algebras $\mathcal Q$ and $\mathcal{IO}$ which define the ``hardware'' of the processors can be included in the specification of the algorithm, or can be taken as the general operator algebra $\mathcal B(H)$ over a Hilbert space.
\item The \emph{input} is provided by setting the edges of the system graph and the inputs $x(v)$ of specific nodes; these settings directly influence the maps $\psi_e$ and $\varphi_v$ which are responsible for communication and local computations within the system, respectively.
\end{itemize}

\begin{definition}
The above described model of a physical system provides a formal characterization of quantum extensions to the \LOCAL model.
\begin{itemize}
\item When no restrictions are made about the state $\omega$, the model is said to be equipped with the $\EN$ extension.

    When state $\omega$ is restricted to be a mixed state separable
    over the local algebras $\mathcal A$, i.e., a state of the form $\sum_i p_i (\omega^i_1\cdots\omega^i_n)$, for some values of probabilities $p_i$, $p_i \geq 0$, $\sum_i p_i = 1$, and some local pure states $\omega^i_n$ over algebra $\mathcal A$, possibly different for each processor, then  the model is said to be equipped with the $\SE$ extension.

    When state $\omega$ is restricted to be of the form $\omega_l \ldots \omega_l$, for some pure state $\omega_l$ over algebra $\mathcal A$, identical for each processor, the system has neither of these extensions.

\item When no restrictions are made about the communication algebra $\mathcal IO$, which may be non-commutative, the system is said to be equipped with the $\QC$ extension.

    When algebra $\mathcal IO$ is restricted to be commutative, communication is understood in the classical sense and the system has no such extension.
\end{itemize}
\end{definition}

Note that the above definition also characterizes the \LOCAL and
\LOCALS models. The fact that such a characterisation is equivalent to
the computational definition (Section~\ref{sec:def}) is
straightforward to prove, taking into account that the considered
initial states are separable, and the proposed evolution cannot create
entanglement since only classical algebras are applied for interaction
between processors (cf.~\cite{Baez87} and \cite{Nielsen99} for
different expositions of related concepts). The evolution on separable
states is then easily simulated by a stochastic process, or
equivalently, a distributed randomized algorithm.

\section{Proof of Theorem: $\LOCALQE \subseteq \PLOCAL$}\label{app:plocal}

\begin{proof}
Consider any quantum distributed algorithm $(\omega, \{\varphi^x\},
\mathcal Q, \mathcal{IO}, M, f)$. For any subset of processors $S
\subseteq V$, we will denote by $\mathcal A^{\otimes S}$ the tensor
product of algebras $\mathcal A$, taken over processors from set $S$,
only. Consider the effect of the evolution maps $\{\varphi_v\}$ and
$\{\psi_e\}$ on an arbitrary operator $A_S \in \mathcal A^{\otimes
S}$. It is clear that $A_S$ is not affected by $\varphi_v$ if $v
\notin S$, and moreover $A_S$ is not affected by $\psi_e$ if $e\notin
\sV_1(S)$. Thus, since these maps encode the input graph $G_x$, after
a single round of evolution we may write $[A_S]_{(1)} = \tau_1 (A_S,
\sV_1(G_x,S))$, where $\tau_1$ denotes some operation dependent only
on the algorithm (independent of $G_x$). Moreover, since maps
$\varphi_v$ are local and maps $\psi_e$ only act on nearest
neighbours, we have $[A_S]_{(1)} \in \mathcal A^{\otimes
\sV_1(S)}$. By applying the evolution procedure for $t$ rounds, we
immediately obtain by induction that $[A_S]_{(t)} = \tau_t (A_S,
\sV_t(G_x,S))$, where $\tau_t$ denotes some operation dependent only
on the algorithm. Since the above holds for any operator $A_S \in
\mathcal A^{\otimes S}$, distributions of results of all measurements
restricted to set of processors $S$ are independent of the input graph
$G_x$, except for the local view $\sV_t(G_x,S)$. This immediately
implies that all solutions which belong to $\LOCALQE[t]$ also belong
to $\PLOCAL[t]$.
\end{proof}

\section{Supplementary Propositions}

\subsection{Properties of problem $\mathscr{P'}$ (selecting an edge of $K_n\setminus{\{e\}}$)}\label{app:kne}

\begin{proposition}
Problem $\mathscr{P'} \notin \LOCAL[0, e^{-1}]$.
\end{proposition}
\begin{proof}
Consider an outcome $\mathscr{O}$ in $\LOCAL[0]$ which solves $\mathscr{P'}$ with some probability $\Pi$. Within $\LOCAL[0]$, the output value $y_i \in\{0,1\}$ of each node $i$ is dependent only on the input label of the node, hence we may assume that with some probability $p_i$ node $i$ returns $0$, and with probability $1-p_i$ returns $1$.
From now on we will only consider the nodes which return $1$ with non-zero probability, i.e. $p_i<1$; w.l.o.g.\ let us suppose that this is the set of nodes $\{1,\ldots,k\}$, for some $k$, $2\leq k\leq n$. W.l.o.g.\ we can assume that $p_1 \leq p_2 \leq \ldots \leq p_k$. It is straightforward to see that the worst-case input $G_x$ for such a procedure is a graph with a missing edge between nodes $1$ and $2$. Hence, we can consider a simple application of Bernoulli's formula for a sequence of $k$ independent random trials $(1,\ldots,k)$ with failure probabilities $(p_1,\ldots,p_k)$, in which the ``winning event'' is the success of exactly two trials, different from the pair $\{1, 2\}$. Denoting $q_i = p_i^{-1} -1$, we have\footnote{When $p_i=0$, all subsequent expressions are well defined through their limit values when $p_i \to 0_+$.}:
\begin{equation}
\label{eq:pi}
\Pi = p_1 \cdots p_k \cdot\Big(\sum_{1\leq i<j\leq k}\!\! q_i q_j - q_1q_2\Big).
\end{equation}
It now suffices to prove that the above expression, treated as a multi-variable function with respect to $k$ and $\{p_1,\ldots,p_k\}$, subject to the constraints $k\in \{2,\ldots,n\}$, $0\leq p_1 \leq p_2 \leq \ldots \leq p_k < 1$, does not achieve a value $\Pi \geq e^{-1}$. This is a technical step which, for the sake of completeness, we perform  below.

First, by directly solving a multi-variable optimization problem for small values of $k$, we establish that $\Pi < e^{-1} \approx 0.36$ for all $k\leq 5$. Indeed, when $k=2$ we verify that $\Pi = 0$, when $k=3$, $\Pi \leq \frac{8}{27} < 0.30$ (and $\Pi = \frac{8}{27}$ is attained when $p_1=p_2=p_3 = \frac{1}{3}$); when $k=4$, $\Pi \leq \frac{5}{16} < 0.32$ (and $\Pi = \frac{5}{16}$ is attained when $p_1=p_2=p_3=p_4=\frac{1}{2}$), when $k=5$, $\Pi \leq (\frac 3 4)^4 < 0.32$ (and $\Pi = (\frac 3 4)^4$ is attained when $p_1=0$ and $p_2=p_3=p_4=p_5=\frac{3}{4}$).

Now, let $k\geq 6$; we will show that it suffices to consider the case when $p_2 = p_3 = \ldots p_{k-1}$. Suppose, to the contrary, that $p_{k-1} > p_{2}$ and that $\Pi \geq e^{-1}$; we will construct an outcome $\mathscr{O'}$ corresponding to some $k'$ and some set of probabilities $\{p_1',\ldots,p_k'\}$, such that either $|\{i : p_i'=p_2'\}| > |\{i : p_i=p_2\}|$ or $k'<k$, and $\mathscr{O'}$ yields a correct solution with probability $\Pi' \geq \Pi$. To achieve this, we put: $k'=k$, $p_i' = p_i$ for all $i<k-1$, $p_{k-1}' = \max\{p_2, p_{k-1} p_k\} < p_{k-1}$, and $p_k' = \frac{p_{k-1}p_k}{p_{k-1}'} > p_k$ (note that either $p_{k-1}'=p_2$, or else $p_k'=1$; in the latter case, we set $k'=k-1$). For convenience of notation we do not sort the values $p_i'$ in non-decreasing order with respect to $i$, but note that we still have $p_1' \leq p_2' \leq p_3',\ldots,p_k'$, so $\Pi'$ is given by an expression analogous to \eqref{eq:pi}. We now introduce some auxiliary notation. For any subset of nodes $S\subseteq \{1,\ldots,k\}$ with $|S|\geq 2$, let $s_0$, $s_1$, and $s_2$ be the probabilities that exactly $0$, $1$, and $2$ of the nodes from $S$, respectively, return a value of $1$ in $\mathscr{O}$ ($s_0'$, $s_1'$, and $s_2'$ are likewise defined for $\mathscr{O'}$; we also extend this notation to letters $x$ for set $X$, and $y$ for set $Y$). Denoting $\gamma_{(S,w)} = \sum_{i \in S} q_i^w$, for $w\in\{1,2\}$, we recall the following simple relations:
\begin{align}
& s_0 = \prod_{i\in S} p_i\nonumber\\
& s_1 = \prod_{i\in S} p_i \cdot \sum_{i\in S} q_i = s_0 \gamma_{(S,1)}\nonumber\\
& s_2 = \prod_{i\in S} p_i \cdot \sum_{i,j\in S,\ i<j}\!\! q_i q_j = \frac{1}{2} s_0 (\gamma_{(S,1)}^2 - \gamma_{(S,2)}) \leq \frac{s_1^2}{2s_0}\label{eq:bound2}\\
& s_0+s_1+s_2 \leq 1\nonumber
\end{align}
Now, let $X = \{1,\ldots,k-2\}$ and $Y=\{k-1,k\}$. By introducing the above relations into formula~\eqref{eq:pi}, we obtain for outcome $\mathscr O$:
\begin{equation}\label{eq:sim}
\Pi = x_0 y_2 + x_1 y_1 + (x_2 - x_0 q_1 q_2) y_0.
\end{equation}
When writing an analogous expression for $\mathscr O'$, we observe that $x_0' = x_0$, $x_1' = x_1$, $x_2' = x_2$, $y_0' = y_0$, $q_1' = q_1$, and $q_2' = q_2$, hence:
$$
\Pi' = x_0 y_2' + x_1 y_1' + (x_2 - x_0 q_1 q_2) y_0.
$$
By subtracting the above expressions we obtain and noting that $y_1 - y_1' = -(y_2 - y_2')$, we obtain:
$$
\Pi' - \Pi = (x_1 - x_0)(y_1' - y_1).
$$
We have that $y_1' - y_1 > 0$ (because $p_{k-1}p_k = p_{k-1}'p_k'$, $p_{k-1}' < p_{k-1}$, and $p_k' > p_k$). Consequently, if $x_1 \geq x_0$, then $\Pi' \geq \Pi$. We will now prove that the opposite case, i.e.~$x_1 < x_0$ is impossible. Indeed, supposing that $x_1 < x_0$, consider the maximum possible value of $y_0$, subject to the following constraints on probabilities $x_0, x_1, x_2, y_0, y_1, y_2$ which are then necessarily fulfilled: $y_0 + y_1 + y_2 =1 $, $x_0 + x_1 + x_2 \leq 1$, $x_1 \leq x_0$, $x_2 \leq \frac12 x_1$ (by~\eqref{eq:bound2}), $x_0 y_2 + x_1 y_1 + x_2 y_0 \geq e^{-1}$ (since the left-hand side is an upper bound on $\Pi$ by~\eqref{eq:sim}, and moreover $\Pi\geq e^{-1}$ by assumption). By solving this optimization problem, we obtain $y_0 \leq 1-e^{-1}$ (with equality obtained for $x_0=1$ and $x_1=x_2=y_1=0$). Hence, since $y_0 = p_{k-1}p_k$, and $p_{k-1} \leq p_k$, we have $p_{k-1}\leq \sqrt{1-e^{-1}} < \frac{4}{5}$. Consequently, for all $i\leq k-2$, $p_i \leq p_{k-1} < \frac 4 5$, and $q_i = p_i^{-1} - 1 > \frac{1}{4}$. But then: $\frac{x_1}{x_0} = \gamma_{(X,1)} = q_1 + \ldots + q_{k-2} > \frac 1 4 (k-2) \geq 1$, a contradiction with the assumption $x_1 < x_0$.

Thus, it only remains to solve the case when $k\geq 6$ and $p_2 = p_3 = \ldots p_{k-1}$. But then by developing expression \eqref{eq:pi}, we have:
$$
\Pi = p_2^{k-4} [(1-p_1)(1-p_k) p_2^2 + (k-3)\cdot(1-p_1)(1-p_2)p_2 p_k + (k-2)\cdot (1-p_k)(1-p_2) p_1 p_2 + {\textstyle{k-2 \choose 2}} (1-p_2)^2 p_1 p_k].
$$
This expression involves only four variables, subject to the constraints: $k\in\{6,7,\ldots,n\}$ and $0 \leq p_1 \leq p_2 \leq p_k \leq 1$, and by solving this problem we indeed verify that always $\Pi < e^{-1}$, which completes the proof.
\end{proof}
We remark that the optimal assignment of probabilities to nodes is as follows: for $n=3$ or $n=4$, we put the uniform distribution, $p_i = \frac{n-2}{n}$. For $n\geq 5$, we always select $1$ as the output for node $1$ ($p_1=0$), and put $p_i = \frac{n-2}{n-1}$ for $2\leq i \leq n$. As $n$ tends to infinity, the probability that the obtained output is correct tends to $e^{-1}$ from below. Clearly, in order to obtain such a solution, the nodes must take advantage of their input labels which contain information about the identifiers and $n$.

\begin{proposition}
Problem $\mathscr{P'} \in \LOCALS[0, 1- O(1/N^2)]$, when considering graphs of order $n\geq N$.
\end{proposition}
\begin{proof}[Proof]
It suffices to observe that the helper variable $h(v)$ can simply encode the output, and be set in a randomized way so as to select all edges of the complete graph with equal probability. For any input, the probability that the solution is correct is then $1 - 1/{n \choose 2}$.
\end{proof}

\subsection{Characterization of the \LOCAL Model through Feasible Outcomes}\label{app:local}

For deterministic algorithms in \LOCAL, the output of any processor $v$ after $t$ rounds is a deterministic function of the identifiers of nodes which are located within its view $\sV_t(G_x, v)$.

\begin{claim}
An outcome $G_x\mapsto\{(y^i, p^i)\}$ can be deterministically
obtained in $t$ rounds in the \LOCAL model if and only if $p^1=1$ and
there exists a function $f$ such that, for every node $v$, $y^1(v) = f
(\sV_t(G_x, v))$.
\end{claim}

\enlargethispage{\baselineskip}

Allowing for randomised algorithms slightly increases the power of the
model. The usual scenario is to allow each node $v$ to flip coins
during the execution of the algorithm, which is in fact equivalent to
picking an arbitrary (possibly large) random positive integer $r(v)$
by each node before the start of execution~\cite{Linial92}. Outcomes
in the \LOCAL model can thus be described by counting the number of
random integer assignments leading to a particular output vector.

\begin{claim}
An outcome $G_x\mapsto\{(y^i, p^i)\}$ belongs to $\LOCAL[t]$ if and
only if there exists a function $f$ such that for some integer $k$ we
have $p^i = \frac{1}{k^n} |\{ r \in \{1,\ldots,k\}^V : \forall v\in V,
y^i(v) = f (\sV_t(G_{(x,r)}, v)) \}|$.
\end{claim}

\end{document}